\documentclass[
    ,final            
  ,numberedheadings 
  ]
  {aipproc}

\layoutstyle{6x9}

\begin{document}

\title{Fundamental physics on natures of the macroscopic vacuum under high intense electromagnetic fields with accelerators }


\author{Kensuke Homma}{
  address={Physical Science, Graduate School of Science, 1-3-1 Kagamiyama Higashi-hiroshima Hiroshima}
}

%

\begin{abstract}
High intense electromagnetic fields can be unique probes to study natures of 
macroscopic vacua by themselves. Combining accelerators with the intense field
can provide more fruitful probes which can neither be achieved by only intense
fields nor only high energy accelerators.
We will overview the natures of vacua which can be accessible via 
intense laser-laser and intense laser-electron interactions. 
In the case of the laser-laser interaction, we propose how to
observe nonlinear QED effects and effects of new fields like 
light scalar and pseudo scalar fields which may contribute to 
a macroscopic nature of our universe such as dark energy. 
In the case of the laser-electron interaction, 
in addition to nonlinear QED effects, we can further discuss the nature of 
accelerating field in the vacuum where we can access physics related with event 
horizons such as Hawking-Unruh radiations. We will introduce a recent 
experimental trial to search for this kind of odd radiations.
\end{abstract}

\maketitle


\section{Introduction}
The nature of the macroscopic vacuum such as Dark Energy (DE) with
the energy density of (2.39meV)${}^4$\cite{DE} is one of the
most mysterious and attractive subjects in fundamental physics. 
However, it is supported by only cosmological observations at present.
If we could probe it in laboratory experiments, we may be able to
unveil its nature even locally. In this respect we would like to consider 
following two possible experimental approaches to this problem, 
both of which require strong electromagnetic fields.
Strong lasers can provide relatively macroscopic and coherent fields 
in the vacuum compared to a colliding point of high energy beams.
If we cross two or more laser fields, we can discuss hidden natures of
the vacuum via the observation of photon-photon interactions.
On the other hand, if we put electrons into the strong field, we would
be able to discuss quantum natures of the vacuum related 
with the event horizon due to the acceleration even in the flat space-time.

The first approach is to attribute DE to hidden new fields with small masses
in the vacuum. The cosmological observation can probe the long distance nature 
of the vacuum with the extremely small couplings to matter, 
i.e. massless gravity exchanges. This is of course the standard way, 
but it is difficult to understand what it is locally after all. 
On the other hand, high energy colliders can probe short distance natures with 
not small coupling via exchange of massive bosons. 
Although the high energy collider environment can provide energies 
to produce massive fields, the relevant mass
scale would not be related with DE whose energy density is order of meV${}^4$. 
In contrast to the two extreme cases, 
the intermediate distance nature via boson exchanges with
meV mass scale and very small coupling might provide a new insight on DE.
Actually there is an argument that if those long-lived light bosons (small
coupling to matter) are mapped on the FRW metric, it may explain the observed 
DE density\cite{deVega}. 
For this approach, we would like to propose to utilize high intense laser-laser 
interactions. Compared to conventional tests of the fifth force at short distances 
between massive bodies, the coherent light-light scattering has advantages
that the process is much less affected by background electromagnetic interactions
and we can further check the polarization dependence of the interaction
to distinguish from the know higher order QED process.
This will be explained in section \ref{Section2}.

The second approach is to consider DE as a sort of field theoretical offset 
energies in the vacuum. In the field theory, quantum fields are treated 
as excitations from a vacuum state which certainly contains 
the zero point energy as the offset energy. 
The usual treatment of the fields on the microscopic Minkowski space-time(MST)
is to take normal ordering (put the annihilation operator to most right)
to avoid the appearance of the zero point energy.
This way practically works in the trivial background geometry like MST. 
However, in the actual universe which must be defined on the dynamically 
evolving curved space-time in general,
the treatment of the offset energy must be seriously reconsidered. 
Thus the natural questions are; what is the role of the zero point energy
in macroscopic vacua and what is the proper treatment on the curved space-time.
The Casimir effect is known to appear as the different number of allowed modes in
between a vacuum with a boundary and the vacuum outside of the boundary.
Although this effect implies the existence of the zero point energy, 
we can not extract modes as experimentally observable field components. 
If we could extract fields directly from the vacuum, it would give us hints 
on how the vacuum state is realized in nature more deeply. 
For this purpose, we may be able to introduce an effective 
event horizon by using electron acceleration by strong laser fields in which
the Hawking-Unruh radiation (HUR) can be expected as radiations from the 
thermal vacuum. For this approach, we would like to introduce a trial experiment
to detect the odd radiations as explained in section \ref{Section3}.

\section{Laser-laser interactions}\label{Section2}
In the laser-laser interaction, 
the dominant contribution is of course from the nonlinear QED effect, 
$i.e.$ real photon - real photon scattering which is expected from
the one loop effective Lagrangian;
\begin{eqnarray}\label{EqQED}
\frac{1}{360}\frac{\alpha^2}{m_e^4}
[4(F_{\mu\nu}F^{\mu\nu})^2+7(F_{\mu\nu}\tilde{F}^{\mu\nu})^2]
\end{eqnarray}
which has not been directly confirmed as the real photon
interaction in the optical wave length range, 
because the expected cross section of $\sim 10^{-42}$b is too small 
to be detected. The smallness of the cross section is due to the short distance 
nature with the internal electron mass scale. It should be stressed 
that this smallness broadens the window for undiscovered fields.
If we could use a strong coherent electromagnetic field, 
we may be able to treat it as a refractive index medium in the vacuum. 
The phase velocity in the linearly polarized electromagnetic field target
(so called crossed-field configuration\cite{Gies} where electric filed 
$\hat{E}$ and magnetic
field $\hat{B}$ are perpendicular with same strengths) is expected to be
\begin{eqnarray}\label{EqPhaseVelocity}
v_{\parallel} = 1 - \frac{8}{45}\frac{\alpha^2}{{m_e}^4}\frac{z_k}{k^2} 
\nonumber \\
v_{\perp}     = 1 - \frac{14}{45}\frac{\alpha^2}{{m_e}^4}\frac{z_k}{k^2}
\end{eqnarray}
where $k$ is wave number of the probe electromagnetic field
and $z_k$ is defined as
\begin{eqnarray}
\frac{z_k}{k^2} = \epsilon^2 (1+(\hat{k}\cdot\hat{n}))^2
\end{eqnarray}
where $\epsilon = E = B$ and $\hat{n} = \hat{B} \times \hat{E}$ 
in the crossed-field condition.

The order of the refractive index change from that of the normal vacuum is
$10^{-11}$ for the energy density $\epsilon^2$ of 1J/$\mu$m${}^3$. 
The refractive index medium would have polarization 
dependence $i.e.$ birefringence nature and the ratio between parallel and 
perpendicular combinations are based on the balance 
between the first and second term in the effective one loop Lagrangian.

On the other hand, if there are hidden fields which may couple to photons 
with not extremely small coupling strength, the birefringence would deviate 
from the expectation in QED. A scalar type of field and a pseudo scalar
type of field can be appeared via the first and second terms in
Eq.(\ref{EqQED}), respectively. 
In addition if scalar and pseudo scalar fields are below meV, 
the long distance nature would enhance the coherent nature of 
the light diffraction compared to higher order QED.
It is worth noting on the sensitivity of the dimensional coupling strength
$g$ of scalar $\phi$ or pseudo scalar field $\sigma$ to two photons
for a given mass scale $m_{\phi,\sigma}$,
if one could probe the QED nonlinear effect.
In the limit of zero momentum exchange, based on the relation
\begin{eqnarray}\label{EqCoupling}
\frac{\alpha^2}{m^4_e} \sim \left( \frac{g}{m_{\phi,\sigma}}\right)^2,
\end{eqnarray}
we can expect to probe down to $g \sim 10^{-9}$GeV${}^{-1}$ 
for $m_{\phi,\sigma} \sim$ meV, while the present experimental limits 
on the model independent axion (pseudo scalar) search with lasers is limited to
$g \sim 10^{-6}$GeV${}^{-1}$\cite{PVLAS}\cite{PVLAS2}.

The experimental key issue is whether we can detect this kind of extremely 
small refractive index change in realistic ways.
In order to increase the amount of the phase velocity shift,
$i.e.$ the energy density of the target laser field, it is essential 
to use a focused laser pulse with an extremely small time duration
with a large energy per pulse. 
To detect the tiny phase velocity shift due to the
high energy density target laser pulse, we need a strong probe laser
to enhance the visibility. However, if we utilize usual interferometer
techniques which assume homogeneous contrast over the probe pulse profile, 
such small refractive index changes would never be detectable.
Instead, here we propose to use the inhomogeneous phase contrast inside 
the probe pulse profile. 
\begin{figure*}[tb]
\includegraphics[width=13.0cm]{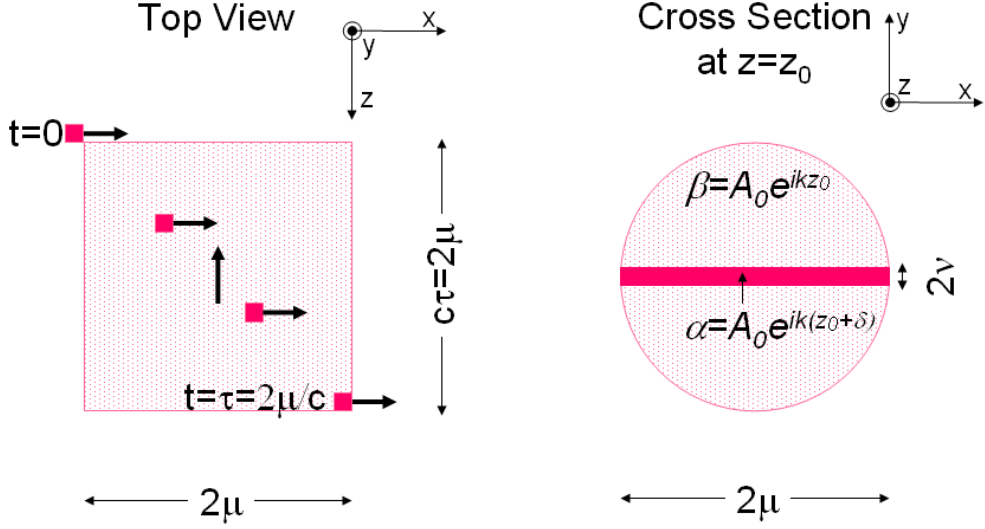}
\caption{
Left: top view of the collision geometry
where a probe laser pulse with a larger profile with the radius of $\mu$
propagating along $z$ is perpendicularly crossed with
a focused target laser pulse which is incident along $x$.
The relative positions of the dense targe laser pulse with respect to the probe
laser pulse in the duration of the crossing time $\tau$ are drawn as a function 
of time $t$ in the laboratory frame.
Right: cross section of the probe pulse at the ejection point $z=z_0$
where the probe laser pulse will contain a trajectory with 
the phase shift along the path of the target laser.
}
\label{Fig1}
\end{figure*}

Suppose a collision geometry as shown in Fig.\ref{Fig1}(left)
where a probe laser pulse with a larger profile with the radius of $\mu$
(less energy density) propagating along $z$ is perpendicularly crossed with 
a focused target laser pulse which is incident along $x$.
The relative positions of the dense target laser pulse with respect
to the probe laser pulse in the duration of the crossing time $\tau$ are 
drawn as a function of time $t$ in the laboratory frame.
In this geometry, after the penetration of the probe laser pulse,
the profile on the x-y plane of the probe laser will contain a trajectory 
with the phase shift along the path of the target laser
as shown in Fig.\ref{Fig1}(right).

Let us approximate that the trajectory has a rectangular 
shape with the size of $2\mu \times 2\nu$.
The inside of the rectangle $rec$ and its outside $\bar{rec}$ are
expressed as follows;
\begin{eqnarray}
rec(\mu, \nu) = \left\{
\begin{array}{ll}
1 & \quad \mbox{for $|x|\le\mu \cap |y|\le\nu$} \\
0 & \quad \mbox{for $|x|>\mu \cup |y|>\nu$}
\end{array}
\right\}
,
\bar{rec}(\mu, \nu) = \left\{
\begin{array}{ll}
0 & \quad \mbox{for $|x|\le\mu \cap |y|\le\nu$} \\
1 & \quad \mbox{for $|x|>\mu \cup |y|>\nu$}.
\end{array}
\right\}
\end{eqnarray}

In order to extract the local phase shift along the target laser trajectory,
measuring the diffraction pattern on a focal plane via
a lens has many advantages.
It is instructive to discuss the feature of the pattern qualitatively.
The intensity pattern obtained on the focal plane 
with a rectangular slit of $2\mu \times 2\nu$ corresponds to 
the Fourier transformation of the slit shape expressed as \cite{SIEGMAN}
\begin{eqnarray}\label{EqSlit}
\left(\frac{\sin(\mu\omega_x)}{\mu\omega_x}\right)^2
\left(\frac{\sin(\nu\omega_y)}{\nu\omega_y}\right)^2
\end{eqnarray}
where $\omega_x=\frac{2\pi}{\lambda f}x$ and $\omega_y=\frac{2\pi}{\lambda f}y$
are spatial frequencies for a given position $(x,y)$ with the focal length $f$
and the wavelength $\lambda$.
In the case of the slit with $\mu >> \nu$,
the rectangle profile on the focal plane becomes orthogonally rotated 
line shape with oscillations because the narrower the slit size, 
the smaller the spatial frequency in that direction.
This is a key feature which makes the visibility of the small phase shift
drastically improve, since the amplitude through the slit is expanded to 
outer region by narrowing the slit size, while the input amplitude pattern 
with a Gaussian shape is confined at the focal point with the smaller waist
as explained below in detail.

Given a Gaussian beam profile of $A_0 e^{-a(x_0^2+y_0^2)}$
as the probe laser pulse, the linearly synthesized amplitude at $z=z_0$
after crossing with the target laser pulse can be expressed as
\begin{equation}
\psi(x_0, y_0) = \alpha rec(\mu,\nu)e^{-a(x_0^2+y_0^2)} + \beta \bar{rec}(\mu,\nu)e^{-a(x_0^2+y_0^2)},
\end{equation}
where $\alpha$ and $\beta$ are the plane wave amplitudes at the ejection point
$z_0$ with the local phase shift $\delta$ caused by the local refractive index 
change due to the nonlinear QED effect and without phase shifts respectively,
which are defined as
\begin{eqnarray}\label{EqPlane}
\alpha = A_0 e^{i(kz_0+\delta)} \nonumber \\
\beta = A_0 e^{ikz_0}.
\end{eqnarray}
The Fourier transformation of the synthesized amplitude on
the focal plane is expressed as
\begin{eqnarray}
F\{\psi(x_0, y_0)\} = \alpha F\{rec(\mu,\nu)e^{-a(x_0^2+y_0^2)}\} +
                       \beta F\{\bar{rec}(\mu,\nu)e^{-a(x_0^2+y_0^2)}\}
\mbox{\hspace{12cm}} \nonumber \\
= (\alpha-\beta) \int^{\mu}_{-\mu}\!\!\int^{\nu}_{-\nu}\!\!dx_0dy_0
  e^{-a(x_0^2+y_0^2)} e^{-i(\omega_x x_0 + \omega_y y_0)}
+ \beta \int^{\infty}_{-\infty}\!\!\int^{\infty}_{-\infty}\!\!dx_0dy_0
  e^{-a(x_0^2+y_0^2)} e^{-i(\omega_y x_0 + \omega_y y_0)}.
\mbox{\hspace{10cm}} \nonumber
\end{eqnarray}
\begin{equation}\label{EqF}
\mbox{\hspace{10cm}}
\end{equation}
Here we introduce a coefficient $C_{sig}$ for the first term containing
the information on how much the phase shift is localized, which is defined as
\begin{eqnarray}\label{EqCsig}
C_{sig}(\omega_x, \omega_y) \equiv 
 \int^{\mu}_{-\mu}\!\!dx_0 e^{-a x_0^2} \cos(\omega_x x_0)
 \int^{\nu}_{-\nu}\!\!dy_0 e^{-a y_0^2} \cos(\omega_y y_0)
\end{eqnarray}
and a coefficient $C_{bkg}$ for the second term which is
a sort of background Gaussian part just to enhance the visibility of 
the phase shift;
\begin{eqnarray}\label{EqCbkg}
C_{bkg}(\omega_x, \omega_y) \equiv
\frac{\pi}{a} e^{-\frac{(\omega_x^2+\omega_y^2)}{4a}},
\end{eqnarray}
where $\int^{\infty}_{-\infty} e^{-at^2} dt = \sqrt{\frac{\pi}{a}}$ is used.
Therefor the Fourier transformation becomes
\begin{eqnarray}\label{EqFfinal}
F\{\psi(x_0, y_0)\} = 
(\alpha-\beta) C_{sig}(\omega_x, \omega_y) + \beta C_{bkg}(\omega_x, \omega_y).
\end{eqnarray}
By substituting Eq.(\ref{EqPlane}), (\ref{EqCsig}) and (\ref{EqCbkg}) 
into Eq.(\ref{EqFfinal}), the intensity pattern at the focal point
can be expressed as
\begin{eqnarray}\label{EqIntensity}
|\psi(\omega_{x},\omega_{y})|^2=(\frac{A_0}{f\lambda})^2
\{ 2C_{sig}(C_{sig}-C_{bkg})(1-\cos\delta)+C^2_{bkg} \}.
\end{eqnarray}
As seen from Eq.(\ref{EqIntensity}), the essence of this method
is that the modulating part due to the phase shift $\delta$ is spatially 
separated from the confined strong Gaussian part $C^2_{bkg}$.

\begin{figure*}[tb]
\includegraphics[width=15.0cm]{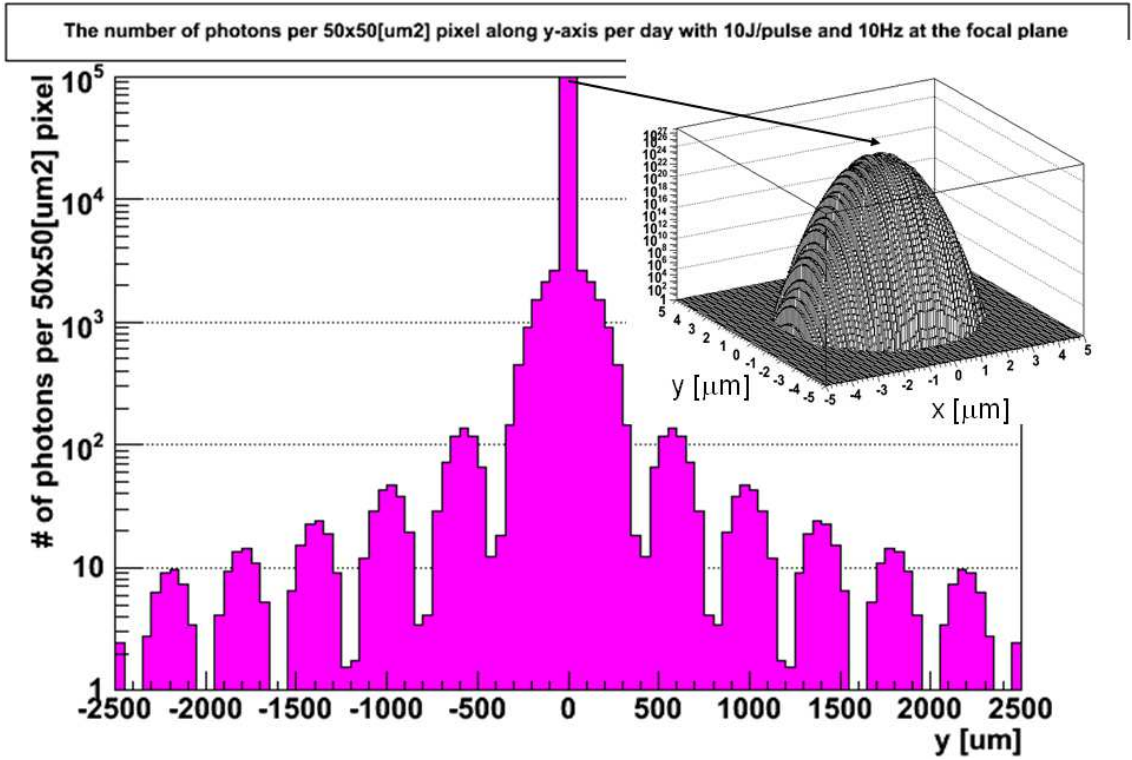}
\caption{
The expected number of photons per $50 \times 50\mu$m${}^2$ pixel
along y-axis at the focal plane calculated by Eq.(\ref{EqIntensity}). 
For this figure, 10~J target laser pulses
are focused in $\pm\nu=10\mu$m with the pulse duration of 1 fs
and 10~J probe laser pulses are focused at the focal plane by
a lens with the focal length $f=5$mm with the pulse duration $\tau=2\mu/c$
where the probe beam waist of $\mu=1.5$mm is assigned as $1\sigma$ 
of the Gauss width. The photon intensity is obtained by accumulating 
over 24 hours with the reputation rate of 10Hz for both pulses.
The lego histogram inside the figure shows the photon intensity 
profile without phase shifts. It is confined within $\pm 5\mu$m.
}
\label{Fig2}
\end{figure*}

Fig.\ref{Fig2} shows the expected number of photons 
per $50\mu$m $\times$ $50\mu$m pixel along the y-axis 
on the focal plane, when 10~J pulses with
the wave length of 800~nm are used for
both target and probe laser pulses integrated over one day data acquisition
with 10~Hz reputation rate. The Lego histogram 
inside the figure shows the photon intensity profile near 
the focal point with $\delta=0$. 
The details of used parameters for the figure are explained in the 
figure caption. If we could sample only the expanded part on the focal plane
which can be spatially separated from the most intense Gaussian
part at the focal point, in principle, 
we can increase the intensity of the probe lights
as much as we like to enhance the visibility of the extremely 
small phase shift by keeping the background part confined at the focal point.
The confinement of the intense part at the focal point is crucial,
since any photo device can not cover a wide dynamic range for intensity 
variations from one to $10^{20}$($\sim 10$J) photons.
In the ideal calculation illustrated in Fig.\ref{Fig2}, we can expect the
sufficient number of photons per camera pixel far from the focal point.
This measurement is possible even with presently available laser systems.
Once the QED nonlinear effect is observed, we can compare the balance
between the first and the second term in Eq.(\ref{EqQED}) by taking
the intensity ratio in different polarization combinations between 
target and probe lasers to investigate effects from undiscovered fields 
in the vacuum.

The verification of the detection principle is under way by 
using an electro-optical crystal with a thin electron beam, 
where the distance of the electron beam to the crystal surface is 
controlled and the controlled refractive
index change along the beam direction can be produced by the electric 
field from the electron beam\cite{NONDESTRUCTIVE}. 
This technique can be applied for the nondestructive measurement of 
slow charged particles as the by-products.

The dominant background source of this measurement would be caused by
the refractive index change by plasmas created in residual gases along the
path of the focused target laser pulse. The refractive index of the plasma in the limit
where collisions between charged particles are negligible is expressed as
\begin{eqnarray}\label{EqNplasma}
N = \sqrt{1-\frac{\omega_p{}^2}{\gamma \omega_0{}^2}}
\end{eqnarray}
where 
      $\omega_0$ is angular frequency of the target laser,
      $\omega_p$ is plasma angular frequency defined as $\sqrt{e^2 n_e / m_e \epsilon_0}$ and 
      $\gamma$ is relativistic factor defined as $\sqrt{1+a_0^2}$ with 
      $a_0=0.85\times10^{-9}\lambda[\mu m]\sqrt{I_0[W/cm^2]}$.
In the low pressure limit of the residual gases, the amount of refractive index change
$\Delta N$ is expressed as $\omega_p{}^2 / 2\gamma \omega_0{}^2$.
Although the refractive index in the plasma becomes smaller, 
the inverted contrast of the phase shift inside the probe pulse still maintains the
line shape along x-axis. Therefore, it would produce the similar characteristic diffraction 
pattern along y-axis on the focal plane eventually. 
In order to reduce this effect, one needs to reduce the electron density $n_e$ in the
residual gases. If one takes $\gamma \sim 1$ as the upper limit of $\Delta N$, 
the air pressure corresponding to the refractive index change of $\sim 10^{-11}$ due to
the nonlinear QED effect with $\sim 1$J/$\mu$m${}^3$ is $\sim 10^{-6}$~Pa.
This pressure can be easily attainable with conventional vacuum pumps.
The collisional frequency due to interactions between electrons and ions
is expected to be $10^8-10^9$s${}^{-1}$ at the critical electron density 
$n_{cr}[cm^{-3}]=1.12\times 10^{21}/\lambda^2[\mu m]$ where $\omega_p$ equals $\omega_0$.
For the target laser pulse duration of fs order, the inverse bremsstrahlung due to the
collisional process in the residual gases is totally negligible operated at $\sim 10^{-6}$~Pa.
Therefore, the dominant background contribution from the residual gas plasma can be suppressed 
with the pressure well below $10^{-6}$~Pa.

Although the reduction of the other instrumental background effects must be performed
in actual experimental situations, the basic advantage of this method 
compared to the birefringence measurement under the static magnetic 
field inside a cavity\cite{PVLAS} is that we can increase the energy density 
of the target laser as much as we like 
without much cares of the damage threshold of cavity optics
given an extremely strong laser source attainable in the very near future.

In addition to this technical aspect, the essential
advantage of this method is that it can provide fundamental 
information on the long range nature of hidden fields. 
Since the target laser is confined in $\pm \nu$ and the QED effect is limited 
in that region due to the short range nature originating from the
electron mass scale, the effective slit size is determined
by the size of the geometrical target laser waist. On the other hand, if a long
range interaction takes place, the effective slit size would become larger.
Based on the argument with Eq.(\ref{EqSlit}), the spatial frequency
on the focal plane in the y-direction would become higher. This causes
the contraction of the fringe pattern along the y-axis.
By combining information on the polarization dependence and the contractive
pattern, in principle, one can argue whether the vacuum contains hidden fields 
with long distance natures or not.

\section{Radiation via event horizon}\label{Section3}
\begin{figure*}[tb]
\includegraphics[width=6.0cm]{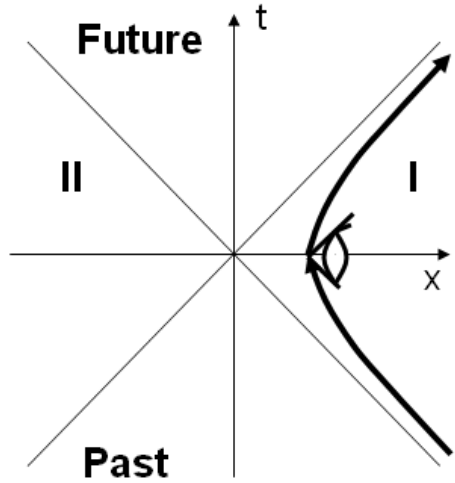}
\caption{
A trajectory of the Rindler observer on the Minkowski space-time.
}
\label{Fig3}
\end{figure*}

Fig.\ref{Fig3} shows a trajectory of an observer with a constant proper 
acceleration $\alpha$ which is referred to as the Rindler observer, 
in a flat Minkowski space-time~(MST) diagram. 
Since the wedge I and II are causally disconnected to that observer, 
there is an effective event horizon in that frame even in the flat MST. 
As an example, suppose a scalar field in MST. 
For an inertial observer, the field can be expressed by a set of mode 
functions and creation-annihilation operators 
However, to the Rindler observer, the field must be dually defined i.e. 
by two sets of mode functions and creation-annihilation operators
in the wedge I and II respectively,
since vacuum states in both wedges are not identical any more. 
This mixing eventually causes the thermal factor in the expectation 
value on the number operator. The corresponding temperature $T$ is 
expressed as
\begin{eqnarray}\label{EqCoupling}
k_B T = \frac{\hbar \alpha}{2\pi c}
\end{eqnarray} where $k_B$ is the Boltzmann constant\cite{Unruh}. 
The radiation from such a thermal bath is referred to as the Unruh radiation.
The essence of the Hawking radiation is same as this which is caused by 
an event horizon due to a black hole in tern. If one is rather interested
in the mechanism of the vacuum thermalization than the black hole phenomenology,
why do we need a real black hole? 
For the lightest charged particle, electron which can be the Rindler observer,
the optical laser field can be an efficient acceleration source
with a macroscopic distance compared to the Compton wavelength of the electron
\cite{Tajima}.

\begin{figure*}[tb]
\includegraphics[width=10.0cm]{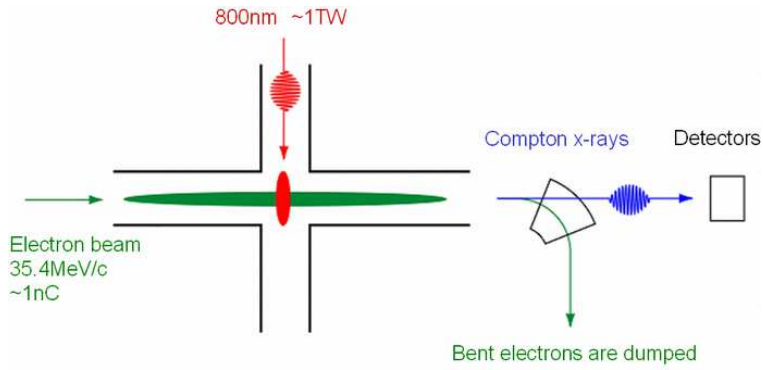}
\caption{
Schematic view of the experimental setup.
}
\label{Fig4}
\end{figure*}

We would like to introduce an experimental trial to search for this kind of 
odd radiations\cite{SUMITOMO}. The set up consists of 35.4MeV electron bunches 
with 1nC and $10^{17}$W/cm${}^2$ Ti:Sa laser pulse with the wavelength of
800~nm in the perpendicularly crossing geometry as illustrated in Fig.\ref{Fig4}.
The laser pulse is linearly polarized and electron bunches are incident parallel
to the direction of the electric field.
At 5m downstream from the crossing point, we put a set of photon 
detectors behind a narrow pin hole to sample X-ray in the limited
acceptance only around the 90 deg crossing angle.
By the pin hole, we can accept only single end-point energy of Compton 
scattering of 14.9keV and guarantee the crossing 
angle is precisely 90 deg by the measured quantity.
We have taken data sets with all four possible combinations between 
laser on(1)/off(0) and electron on(1)/off(0) successively in 10 Hz 
reputation rate. Those combinations will be denoted as
$l_1\&e_1$, $l_1\&e_0$, $l_0\&e_1$ and $l_0\&e_0$ for short.

\begin{figure*}[tb]
\includegraphics[width=11.0cm]{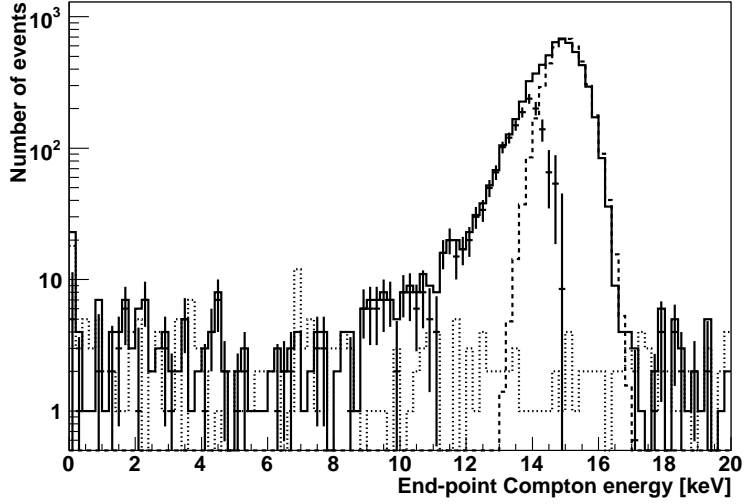}
\caption{
Energy spectra of the end-point region of Compton scatterings 
behind the narrow square slit with the size of $1mm \times 1mm$
at the 5m downstream from the crossing point,
where the solid line shows the spectrum with $l_1 \& e_1$ (signal + beam halo),
the dotted line shows the spectrum with $l_0 \& e_1$ (beam halo),
the dashed line shows the fit to only the right side of the peak, and
the points with error bars show a spectrum by subtracting the fit result
from the spectrum with the solid line.
}
\label{Fig5}
\end{figure*}

Fig.\ref{Fig5} shows the line energy spectrum of the single end-point Compton 
ray in log scale. The solid line is the energy spectrum obtained in $l_1\&e_1$
and the dotted line is in $l_0\&e_1$.
The measured peak exactly corresponds to the Compton end-point energy 
by taking the energy resolution of the X-ray detector into account.

In order to search for another type of radiations with much lower energies,
we tried to measure the radiation in visible-UV wavelength region.
For this purpose, after we guaranteed the perpendicular crossing geometry, 
we put a filter to reduce stray lights from the upstream laser, 
a mirror and a photomultiplier tube~(PMT) with the timing resolution below 1~ns
in front of the X-ray detector so that the mirror reflects 
visible-UV range lights from the crossing point to PMT which is only sensitive 
to visible-UV range, if such lights are emitted from the crossing point.

\begin{figure*}[tb]
\includegraphics[width=10.0cm]{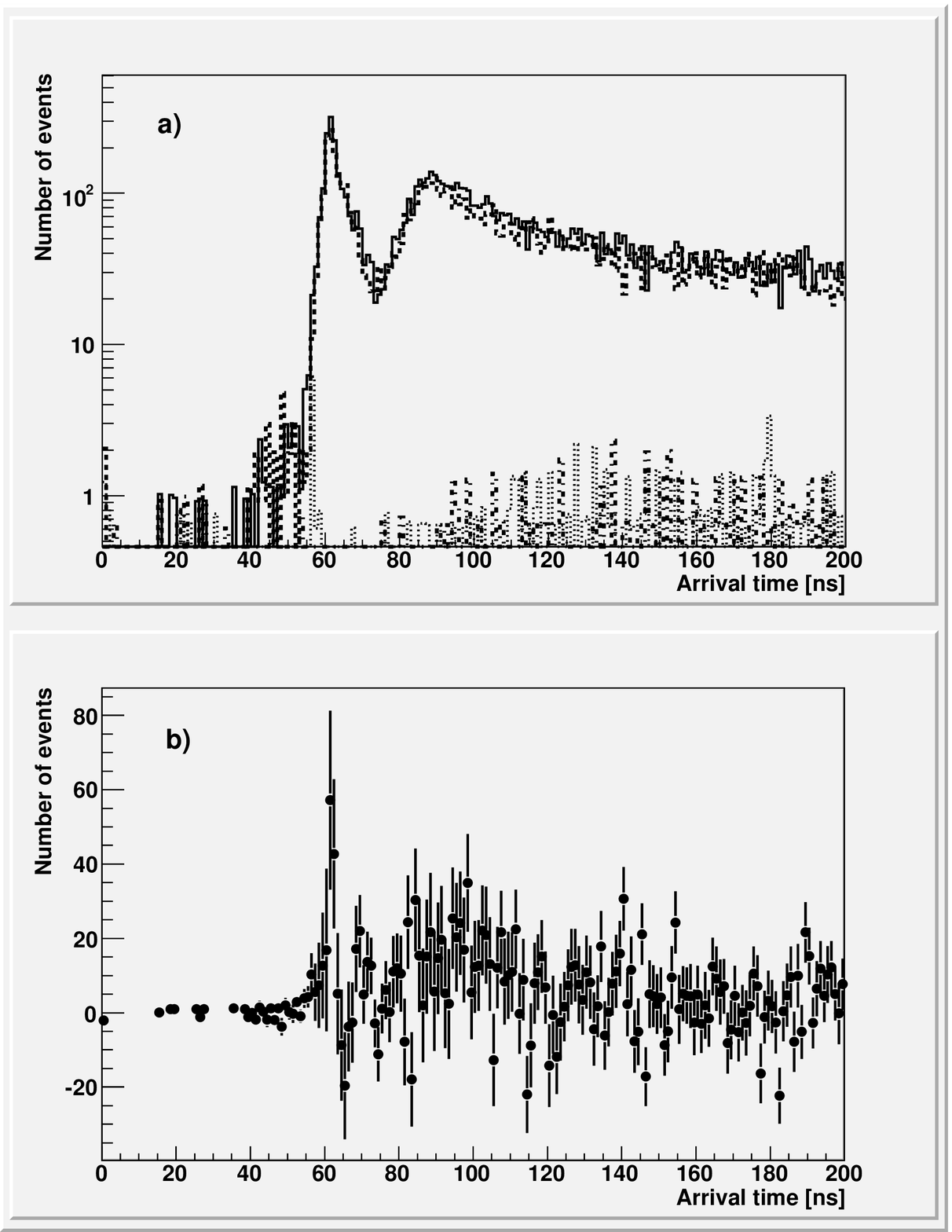}
\caption{
a) Time of flight distributions of visible-UV rays, 
where the solid line is the case 1)
$l_1 \& e_1$ (signal + all backgrounds), the dashed line is the case 2)
$l_1 \& e_0$ (background mainly from laser leaks from the upstream 
after laser dumps), the dotted line is the case 3) $l_0 \& e_1$ 
(backgrounds mainly from electron beam halos), and the dash-dotted line is 
the case 4) $l_0 \& e_0$ (only electrical backgrounds).
b) Difference of the time of flight distributions
between case 1) $l_1 \& e_1$ and 2) $l_1 \& e_0$. 
}
\label{Fig6}
\end{figure*}

Fig.\ref{Fig6}~a) shows the time of flight distributions between the reference 
clock synchronized with the beam bunches and the hit time in PMT
located at 5m down stream from the crossing point.
The solid line is the case 1) $l_1 \& e_1$ (signal + all backgrounds), 
the dashed line is the case 2) $l_1 \& e_0$ (background mainly from 
laser leaks from the upstream after laser dumps), 
the dotted line is the case 3) $l_0 \& e_1$ (backgrounds mainly from 
electron beam halos), and the dash-dotted line is
the case 4) $l_0 \& e_0$ (only electrical backgrounds).
The reason why the case 2) sees the peak like structures is 
that the laser beam dump was located within 30~cm from the crossing point and 
the reflection from the dump reaches the PMT even after the cut off filter 
to 800~nm in front of the mirror and the reflected lights further go back and 
forth between the dump and the mirror. 
Therefore the primary peak position corresponds to a time  when a light 
from the crossing point reaches PMT in the down stream. 
If there are additional light emissions from the crossing point 
over the laser dumping yield, one should see the enhanced yield 
at this narrow timing window. 
Fig.\ref{Fig6}~b) shows the hit time distribution after subtracting 
the case 2) from the case 1). The distribution suggests
a possible enhancement at the proper hit timing in visible to UV range. 

The reason why we choose the 90 degree crossing geometry was 
because the classical Larmor radiation effect is maximally suppressed 
in that geometry. The numerical simulation where electrons are wiggled 
inside a linearly polarized laser pulse by assuming only the Lorentz 
force without back reactions gives us an estimate of $10^{-5}$ visible 
rays for the experimentally obtained event statistics of 1.5 K crossing.
Therefore the Larmor radiation is sufficiently suppressed. The other possible 
emission sources like synchrotron radiations by the bent electrons,
linear/non-linear Compton rays and Cherenkov radiations due to the
nonlinear QED effect\cite{Cherenkov} can not explain the visible-UV enhancement.

As an interesting interpretation, the Unruh radiation can be considered here. 
For the given electric field strength of $\sim$TV/m, 
the Unruh temperature corresponds to $\sim 0.06$~eV in the instantaneous rest 
frame of the incident electron and the blue shifted energy in the inertial
(laboratory) frame becomes $\sim 10$ eV which is consistent with 
the visible to UV enhancement. 
Actually we have put the detector system in the sweet spot where the Unruh yield
over the Larmor radiation yield is maximized. In order to verify the radiation 
from the thermal vacuum, one needs more distinct signatures other than 
the thermal spectrum. Since it is known that a successive absorption and 
emission in the Rindler frame which must keep the memory of the 
conserved quanta in the vacuum, can be interpreted as a pair radiation 
in the inertial frame\cite{Unruh2}\cite{Schutzhold},
more promising signature is the strong correlation in both momentum 
and angular momentum in the pair photons compared to the familiar 
Bremsstrahlung process where such correlations are not expected. 
We are planning this kind of correlation measurements 
in addition to the energy measurement in the near future.

\section{Future prospects}
The direct verification of the nonlinear QED effect would be
possible in an ideal experimental setup where a phase velocity shift along a
trajectory of a strong target laser pulse can be measured as
a characteristic diffraction pattern of a probe laser on a focal
plane via a lens even with currently available laser systems.
Once we could observe it, it would be very interesting to check 
the polarization dependence of the diffraction intensity and the spatial
frequency of the diffraction pattern, which can be a model independent test 
to investigate whether vacuum contains
hidden light (long range) scalar/pseudo scalar fields or not. 

Acceleration of electrons with an electric field by a strong laser pulse
can provide a good test ground of radiations via the event horizon introduced 
in Minkowski vacuum. A trial experiment may suggest possible odd radiations.
The correlation measurement would unveil the nature of the odd radiations.

\begin{theacknowledgments}
This work is supported by a Grant-in-Aid for scientific Research 
no. 18654047 and no. 18684009 from the Ministry of 
Education, Culture, Sports, Science, and Technology (MEXT).
\end{theacknowledgments}

\bibliographystyle{aipproc}   
\bibliography{refbib}

\end{document}